\documentclass[12pt]{article}

\usepackage{epsfig}

\newcommand{\be}{\begin{eqnarray}}
\newcommand{\ee}{\end{eqnarray}}
\newcommand{\nn}{\nonumber}
\renewcommand{\d}{\mathrm{d}}
\renewcommand{\i}{\mathrm{i}}

\newcommand{\feynmandag}{\slash\hspace{-0.55em}}
\newcommand{\alphas}{\alpha_{\mathrm{s}}}
\newcommand{\xBj}{x_{\mathrm{Bj}}}
\newcommand{\mDs}{m_{D_s}}
\newcommand{\mc}{m_c}

\begin{document}
\begin{center}
{\LARGE Neutrino induced hard exclusive $D_s$ production}\\[0.8cm]
{\large B.~Lehmann-Dronke and A.~Sch\"afer}\\[0.5cm]
{\footnotesize Institut f\"ur Theoretische Physik,
Universit\"at Regensburg, 93040 Regensburg, Germany}
\end{center}
\vspace{0.5cm}
\begin{abstract}
Motivated by the possibility to use high intensity neutrino beams for
neutrino--nucleon scattering experiments
we analyze charged current induced exclusive meson
production within the framework of generalized parton distributions.
The cross section for hard exclusive $D_s$ production is estimated in
this formalism to leading order in QCD.
The integrated cross section proves to be sizable.
It is shown that the considered process is well suited to provide
novel information on the gluon structure of nucleons, which is
contained in the generalized gluon parton distribution.
\end{abstract}
The factorization theorem \cite{factor,Rad} states that up to power
suppressed terms every contribution to the amplitude
for hard exclusive meson production can be written as a
convolution of a generalized parton distribution (GPD), a distribution
amplitude, and a hard part. This has recently been applied for the
investigation of electroproduction of single light mesons \cite{mesonprod}
and meson pairs \cite{maxben}.  The fact that in the near future also high
intensity neutrino beams might be available for lepton--nucleon
scattering experiments \cite{nufact,nudis} motivates the present study
extending the formalism to charged current induced processes.

\begin{figure}
\centering
\epsfig{file=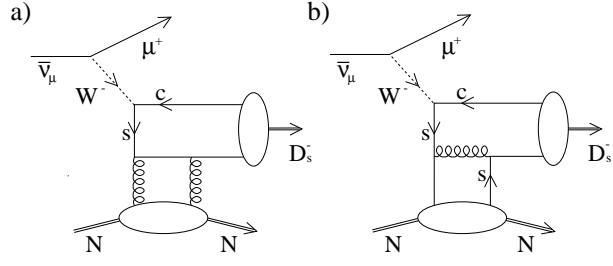, width=8cm}
\caption{Two of the five contributing diagrams.}
\label{graph}
\end{figure}

We consider the example
\be
\bar\nu_{\mu} + N \to \mu^+ + N + D_s^-\,.
\label{process}
\ee
Similar processes have already been subject of experimental studies
\cite{csprod}. For the analysis presented here, however, $Q^2$ has to
be large compared to $-t$ and the squared masses of the involved
particles. The leading order amplitude is given by the sum of three
diagrams involving a gluon GPD (Fig.~\ref{graph}\,a
and diagrams obtained by an interchange of vertices)
and two diagrams with a contribution of the (polarized and unpolarized)
strange quark GPD (Fig.~\ref{graph}\,b plus one
diagram with changed order of vertices).
It is important to note that the only dependence
on quark distributions in the nucleon comes in via the strange
sea. Therefore the considered process is a good probe for the
gluon GPD, which dominates the amplitude.
To leading order in the strong coupling constant $\alphas$ and neglecting
terms of the order $\mDs^4/(Q^2+\mDs^2)^2$ where $\mDs$ is the mass of
the $D_s$ meson the amplitude for the subprocess
\be
W^{-^\ast}_L(q) + N(p)\to D_s^-(q')+N(p')
\ee
is given by
\be
T&=&\langle N(p'), D_s^-(q')| J^{\mathrm{charged}}\cdot
\varepsilon_L|N(p)\rangle
\nn\\
&=&-\frac{e 4 \pi \alphas}{2\sqrt{2}\sin\theta_W}\frac{1}{Q}
\int_0^1 \d z\,\frac{\Phi_{D_s^-}(z)}{z} \int_{-1}^1 \d\tau
\nn\\
&&\times\Biggl\{
\frac{T_F}{N_c}
F_G(\tau,\xi,t)
\biggl[\frac{1}{\tau+\xi-\i\epsilon}+
\frac{1}{\tau-\xi+\i\epsilon}
\biggr]
\nn\\
&&\;\;\times
\Biggl[\frac{1}{(1-z)}+\frac{1}{2}\frac{\mDs^2}{Q^2+\mDs^2}
\frac{1}{(1-z)(\tau+\xi)-\frac{\mDs^2}{Q^2+\mDs^2}2\xi
(1-z-\frac{\mc^2}{\mDs^2})-\i\epsilon}
\nn\\
&&\;\;\;\;\times
\frac{-\tau^2(1-z)^2-\xi^2(1+z(4-8\frac{\mc^2}{\mDs^2})-
z^2(9-4\frac{\mc^2}{\mDs^2})+4z^3)}
{(1-z)(\tau-\xi)+\frac{\mDs^2}{Q^2+\mDs^2}2\xi
(1-z-\frac{\mc^2}{\mDs^2})+\i\epsilon}
\Biggr]
\nn\\
&&\;+
\frac{\mDs^2}{Q^2+\mDs^2}\frac{T_F}{N_c}
\tilde F_G(\tau,\xi,t)
\biggl[\frac{1}{\tau+\xi-\i\epsilon}+
\frac{1}{\tau-\xi+\i\epsilon}
\biggr]
\nn\\
&&\;\;\times
\frac{1}{[1-z-\frac{\mDs^2}{Q^2+\mDs^2}(1-z-\frac{\mc^2}{\mDs^2})]
[(1-z)(\tau+\xi)-\frac{\mDs^2}{Q^2+\mDs^2}2\xi
(1-z-\frac{\mc^2}{\mDs^2})-\i\epsilon]}
\nn\\
&&\;\;\times
\frac{-\tau\xi(1-z)(1-3z+z^2)}
{(1-z)(\tau-\xi)+\frac{\mDs^2}{Q^2+\mDs^2}2\xi
(1-z-\frac{\mc^2}{\mDs^2})+\i\epsilon}
\nn\\
&&\;+
\frac{C_F}{N_c}
\Bigl[F_s(\tau,\xi,t)-\tilde F_s(\tau,\xi,t)\Bigr]
\frac{1}{(\tau+\xi-\i\epsilon)}
\nn\\
&&\;\;\times
\Biggl[1+\frac{\mDs^2}{Q^2+\mDs^2}\Biggl(-\frac{1}{2}+
\frac{\xi}{\tau+\xi-2\xi\frac{\mDs^2-\mc^2}{Q^2+\mDs^2}-\i\epsilon}\Biggr)
\Biggr]
\Biggr\}\,,
\ee
where the skewedness parameter $\xi$ is related to the Bjorken
variable by $\xi=\xBj(Q^2+\mDs^2)/(2Q^2-\xBj(Q^2+\mDs^2))$.
The mass of the charm quark
$\mc$ is understood as pole mass, in the following a value of
$\mc=1.5\,\mathrm{GeV}$ is used. For the GPD's $F_G(\tau,\xi,t)$,
$\tilde F_G(\tau,\xi,t)$, $F_s(\tau,\xi,t)$, and $\tilde
F_s(\tau,\xi,t)$ the notation of Ref.~\cite{Ji} is used and the
distribution amplitude for the $D_s^-$ meson is defined by
\be
\Phi_{D_s^-}(z)&=&
\int\frac{\d\lambda}{2\pi} e^{-\i\lambda z(q'\cdot\tilde n)}
\langle D_s^-(q')|
T\big[\bar\psi^s(\lambda\tilde n)\gamma_5\feynmandag\tilde n
\psi^c(0)\big]
|0\rangle\;.
\ee
The differential cross section for leptoproduction is given
in terms of this amplitude by
\be
\frac{\d\sigma}{\d\xBj\d Q^2\d t}
=
\frac{e^2}{2(4\pi)^3\sin^2\theta_W}\frac{\xBj}{Q^2(Q^2+M_W^2)^2}
\biggl(1-\frac{Q^2}{2\xBj\,p\cdot l}\biggr)
\sum_{s'}|T|^2\;,
\ee
where $l$ is the neutrino momentum.

For an numerical estimate of the cross section we model the
distribution amplitude following \cite{Bauer} as
\be
\Phi_{D_s^-}(z)=N_{D_s}\sqrt{z(1-z)}
\exp\bigg[-\frac{m_{D_s}^2}{2\omega^2}z^2\bigg]
\label{DA}
\ee
taking for the parameter $\omega$ the value $\omega=1.38\,\mathrm{GeV}$
obtained in \cite{omega} as the best fit for the $D$ meson.
The normalization constant $N_{D_s}$ has to be chosen such that the sum rule
\be
\int_0^1 \d z\,\Phi_{D_s^-}(z)=f_{D_s}
\ee
is satisfied, where we adopt for the decay constant $f_{D_s}$ the
value $f_{D_s}=270\,\mathrm{MeV}$ as an average of the results
obtained so far in lattice calculations \cite{fDs}, see also \cite{Rich} for
an earlier review.

The gluon and quark GPD's are parameterized as in \cite{maxben}
combining Radyushkin's model \cite{Rad,Rad99} with the
parameterizations of the usual (forward) parton distributions of
Refs.~\cite{MRS} (MRS (A')) and \cite{GS} (Gluon A (NLO)).
For the $t$-dependence of the GPD's we adopt the factorized ansatz
$F(\tau,\xi,t)=F(\tau,\xi,0) F_\theta(t)/F_\theta(0)$ and use the
parameterization of \cite{BGMS93} for the gluon form factor
$F_\theta$. For the strong coupling constant the two loop result is
taken with $N_f=4$ and $\Lambda^{(4)}_{\mathrm{QCD}}=250\,\mathrm{MeV}$.

\begin{figure}
\centerline{\epsfig{file=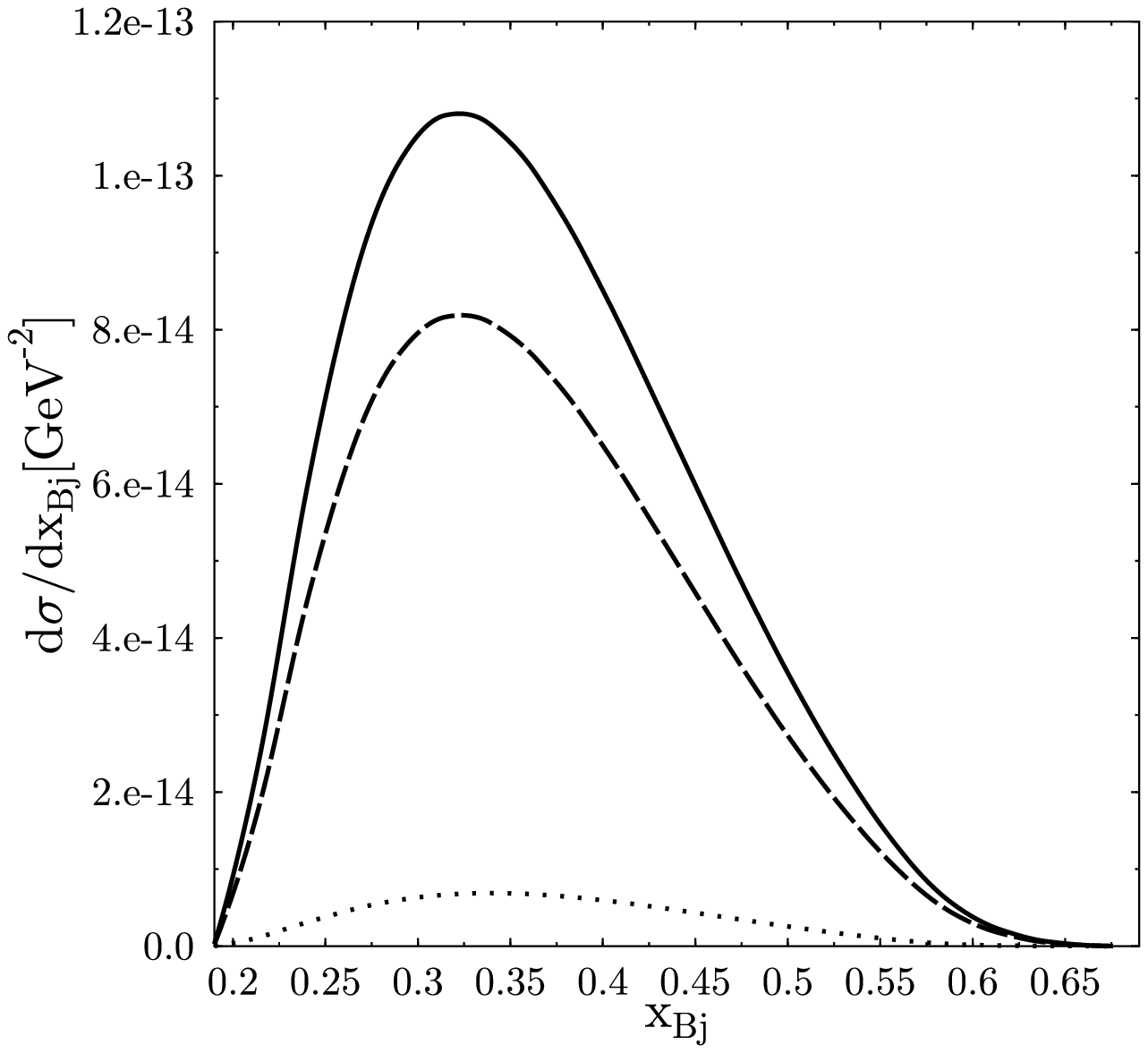, width=7.5cm}
            \epsfig{file=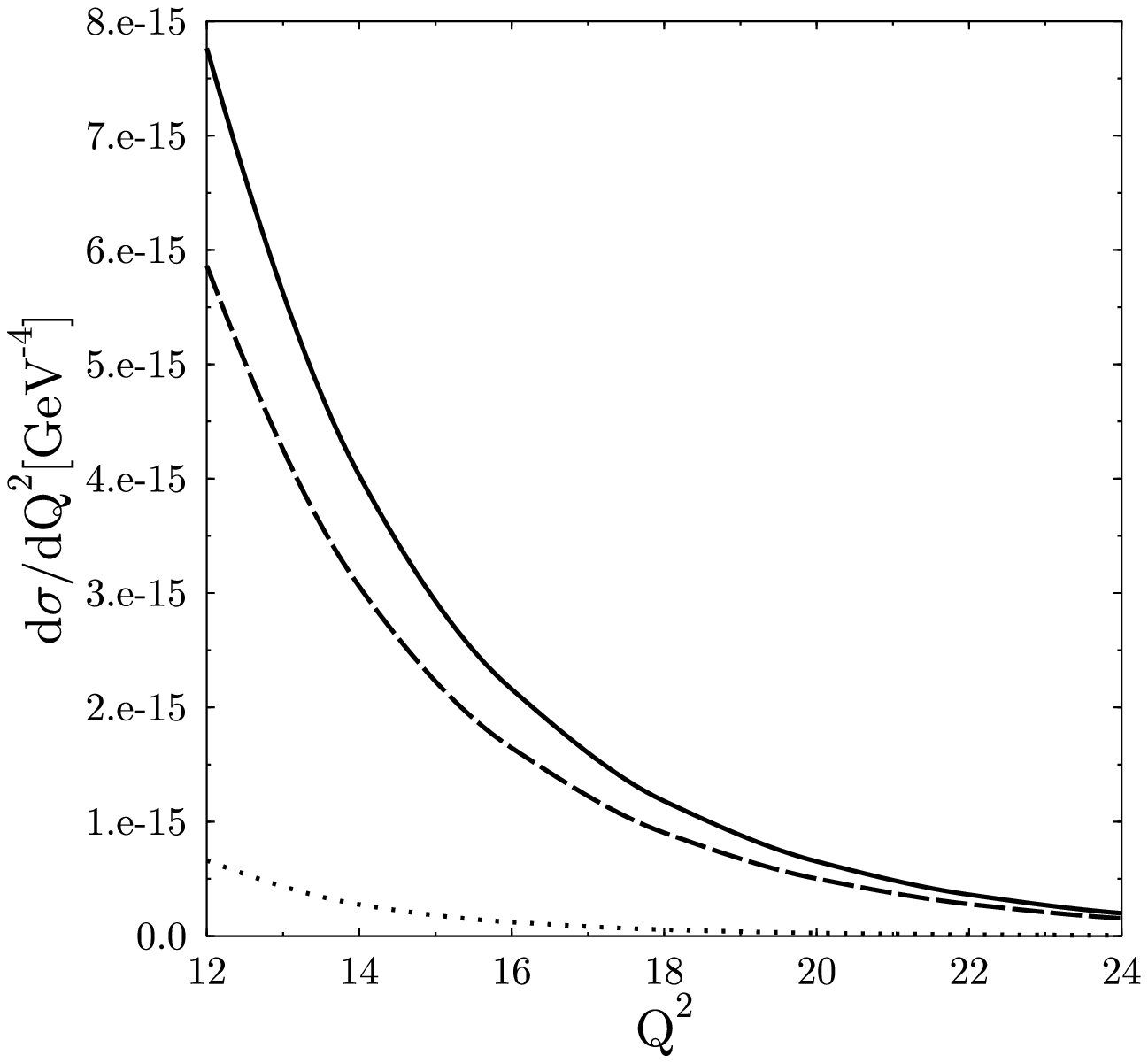, width=7.5cm}}
\caption{The differential cross section for exclusive $D_s^-$ production
  as a function of $\xBj$ or $Q^2$ respectively. The dashed lines show
  the contribution stemming from the gluon GPD $F(\tau,\xi,t)$
  alone. The small contribution of the polarized gluon GPD
  $\tilde F(\tau,\xi,t)$ multiplied with a factor 100
  is plotted with dotted lines.}
\label{result1}
\end{figure}
\begin{figure}
\centerline{\epsfig{file=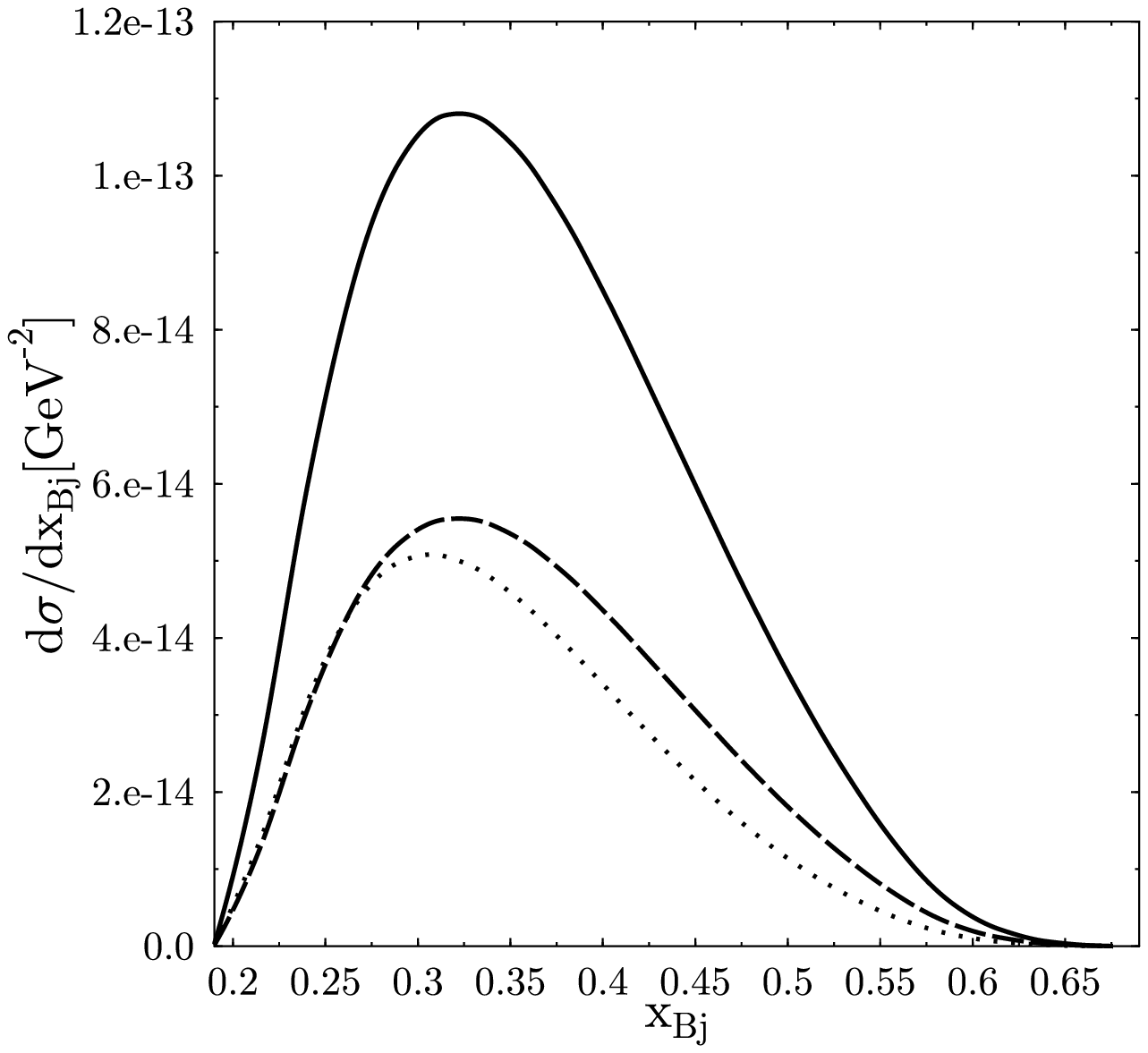, width=7.5cm}
            \epsfig{file=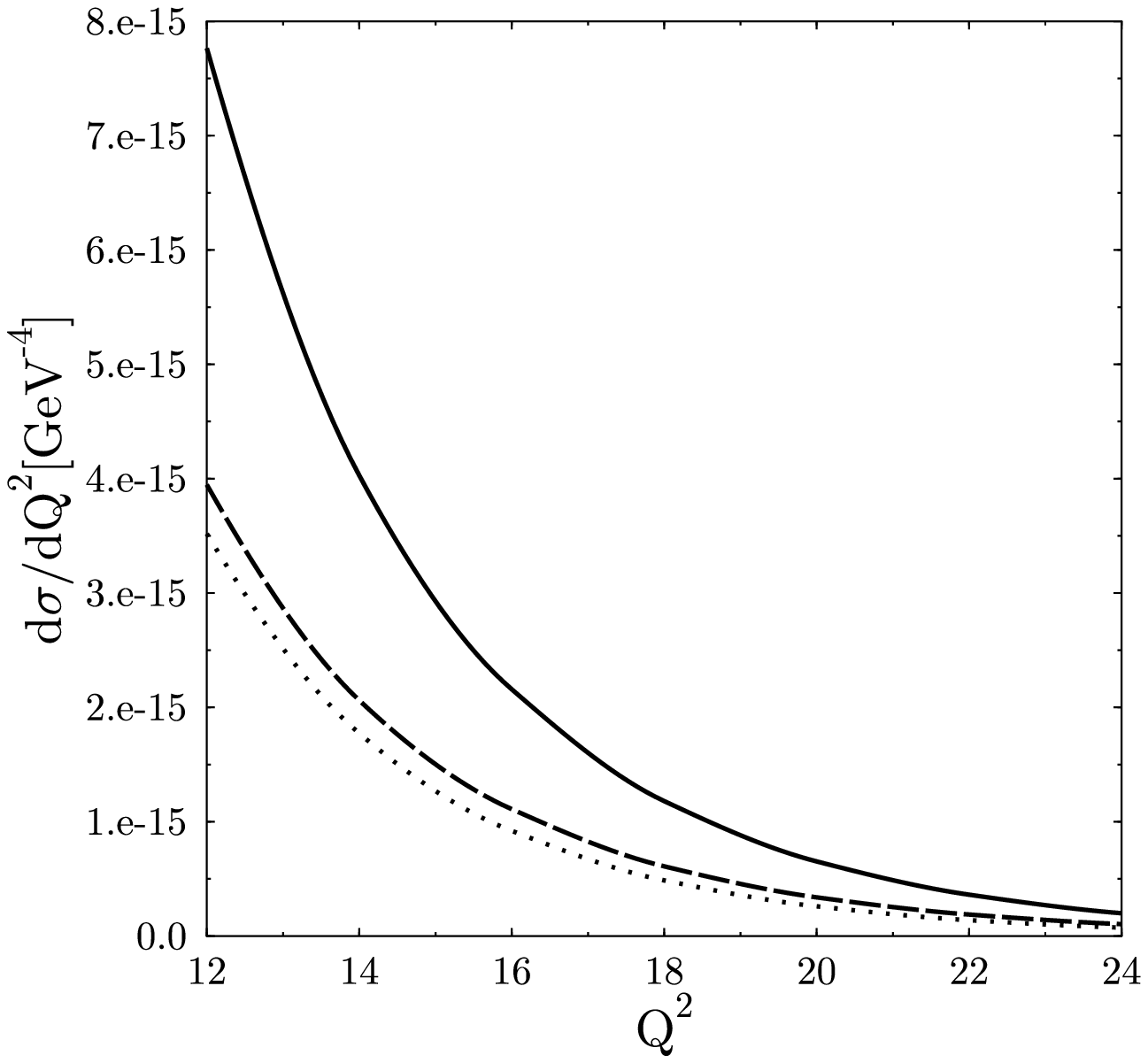, width=7.5cm}}
\caption{The differential cross sections from Fig.~\ref{result1}
  compared with the results obtained for the asymptotic form of the
  distribution amplitude $\Phi_{D_s^-}$ (dashed lines) and by modeling 
  the GPD's by their forward limit (dotted lines).}
\label{result2}
\end{figure}
\begin{figure}
\centerline{\epsfig{file=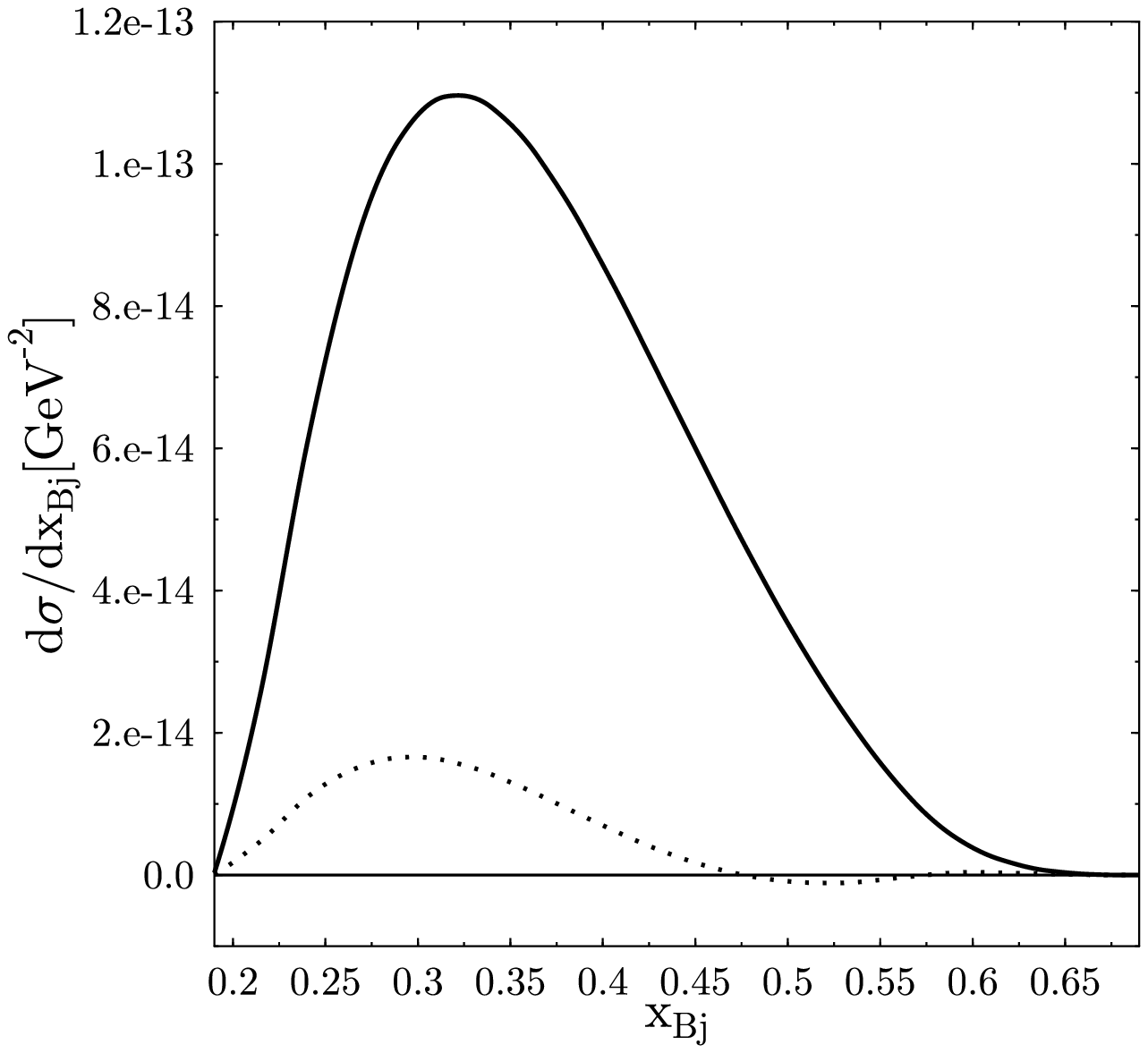, width=7.5cm}
            \epsfig{file=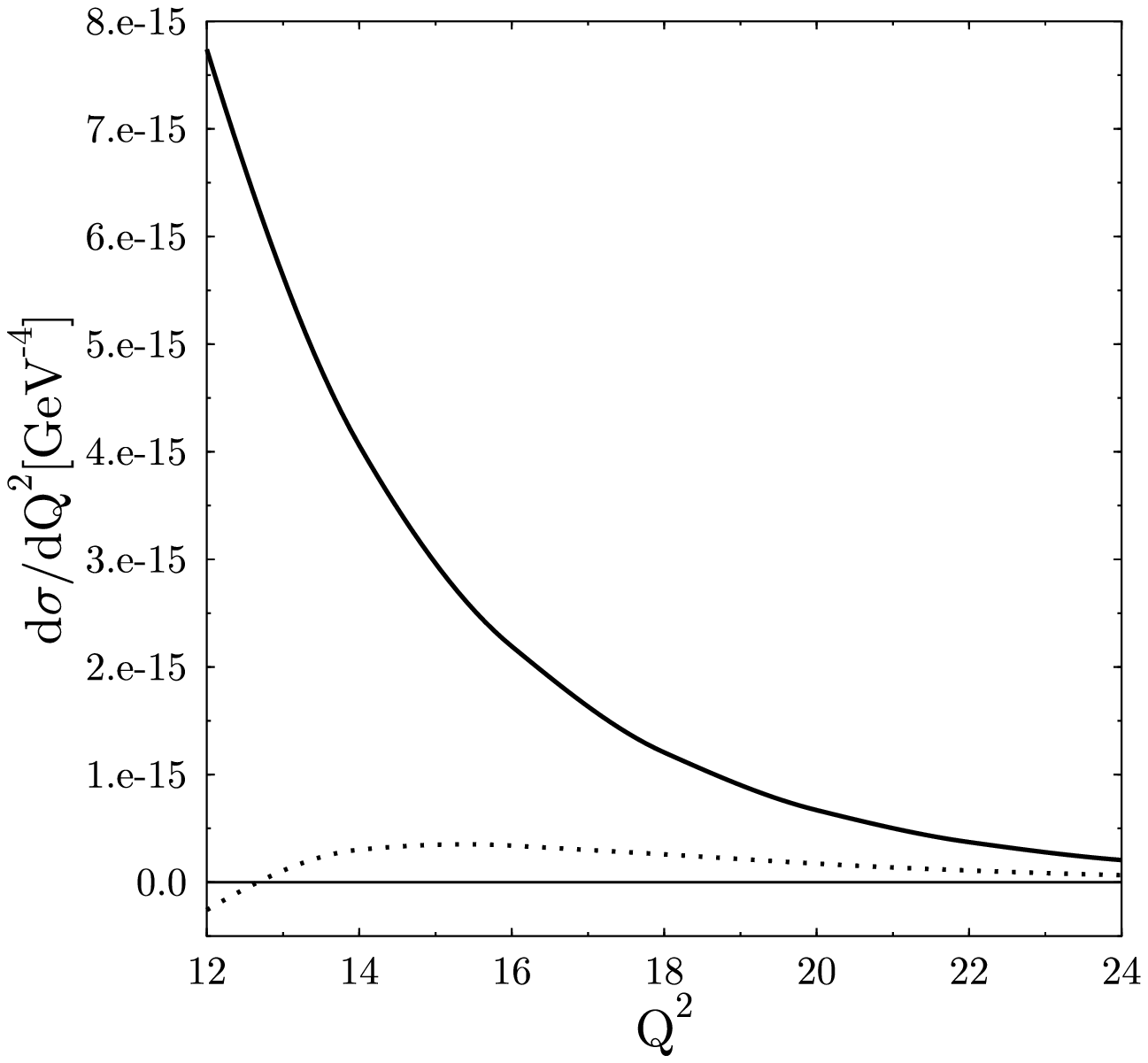, width=7.5cm}}
\caption{The result without mass corrections of the order
  $\mDs^2/Q^2$. The very small difference to the complete result
  multiplied with a factor 10 is plotted with dotted lines.}
\label{result3}
\end{figure}

Figure \ref{result1} shows the results obtained for the differential
cross sections $\d\sigma/\d\xBj$ and $\d\sigma/\d Q^2$ where
$t=(p-p')^2$ has been integrated over the interval
$t_{\mathrm{min}}=m_N^2\xi^2/(1-\xi^2)<-t<2\,\mathrm{GeV}^2$ and the
neutrino energy has been chosen as $E_{\bar\nu}=34\,\mathrm{GeV}$. For the
plot of the $\xBj$-dependence $Q^2$ has been integrated from
$12\,\mathrm{GeV}^2$ to the upper bound given by the constraint
$y<1$, with $y:=p\cdot q/p\cdot l=Q^2/(2\xBj p\cdot
l)$. The plot of $\d\sigma/\d Q^2$ is based on the $\xBj$-dependent
cross section integrated between $\xBj=0.19$ and $\xBj=0.69$ taking into
account the same kinematical constraints.
The dashed lines are obtained neglecting the contribution of the
strange quark GPD and the polarized gluon GPD proving the dominance
of the contribution of $F_G(\tau,\xi,t)$. The negligible small
contribution of the polarized gluon GPD multiplied with a factor 100
is plotted with dotted lines.

To illustrate the dependence of the cross section on the shape of the
$D_s$ distribution amplitude and the GPD's we show in Fig.~\ref{result2}
the results obtained by modeling $\Phi_{D_s^-}$ using its asymptotic form
$\Phi_{D_s^-}(z)=6f_{D_s}z(1-z)$ instead of the parameterization of
Eq.~(\ref{DA}) (dashed lines) and alternatively by replacing the models
for the GPD's $F(\tau,\xi,0)$ by their forward limit $F(\tau,\xi,0)$
(dotted lines). The latter choice corresponds to the approximation of the
GPD's for $t=0$ by usual parton distributions. For comparison also the
original result is plotted again in Fig.~\ref{result2}.

In Fig.~\ref{result3} we finally show the result that is obtained if
all corrections of the order $\mDs^2/Q^2$ in the amplitude are neglected.
The very small difference to the complete result multiplied with a
factor 10 is plotted with dotted lines. The smallness of these
corrections justifies the approach based on the factorization theorem,
which is valid only up to terms of the order $\mu^2/Q^2$
with $\mu$ being a typical mass scale of the process.

Integrating all variables $Q^2$, $\xBj$, and $t$ over the
kinematical region specified above gives a value for the total cross section of
$\sigma=2.45\times 10^{-14}\,\mathrm{GeV}^{-2}
=9.5\times 10^{-6}\,\mathrm{pb}$.
Even given this small value the huge integrated luminosities of $\int
L\d t>10^{45}\,\mathrm{cm}^{-2}=10^{9}\,\mathrm{pb}^{-1}$ available
at a neutrino factory \cite{nufact} would lead to a sizable number of
events of the order of magnitude $10^4$.
Larger total cross sections can be obtained for higher neutrino
energies because of the larger available kinematical region.

Uncertainties of the rough estimate presented here
result from the lack of knowledge about the exact form of the
meson distribution amplitude as illustrated in Fig.~\ref{result2}. Also
the predictions for the value of $f_{D_s}$ differ by about $20\%$.
Up to now the experimental uncertainty for $f_{D_s}$ is even larger
\cite{fDsexp}. It is worth noting, however, that experiments of the
kind discussed in the present article can provide much more precise 
information on the $D_s$ decay constant independently of
the exact cross section for the process (\ref{process}).
As discussed in \cite{Migliozzi} the relatively high production rate
of $D_s$ mesons allows to determine $f_{D_s}$ by
measuring the $D_s$ branching ratios and its total width.

From the experimental point of view a difficulty arises from the fact
that all three particles in the final channel need to be detected
because the neutrino beam energy is not sharp and therefore
an exclusive event can not be identified by reconstructing the
momentum of an undetected particle via a missing momentum analysis.

It has been shown that the high intensity neutrino beams available at
neutrino factories allow to study hard exclusive meson production also
for weak interaction induced reactions opening a new possibility to
study the nucleon structure and allowing to better test the theory of
these exclusive processes. The production of charmed strange mesons
proves to be of particular interest due to its high sensitivity to
the gluon GPD's.

The authors thank S.I.~Alekhin, V.M.~Braun, A. Freund, A.~Lenz, M.L.~Mangano,
P.~Migliozzi, S.~Schaefer, E.~Stein, and C. Weiss for helpful discussions.
This work has been supported by the Studienstiftung des deutschen Volkes.


\begin{thebibliography}{99}
\bibitem{factor}
J.~Collins, L.~Frankfurt, and M.~Strikman,
Phys.\ Rev.\ D {\bf 56}, 2982 (1997).
\bibitem{Rad}
A.V.~Radyushkin, Phys.\ Rev.\ D {\bf 56}, 5524 (1997).
\bibitem{mesonprod}
S.J.~Brodsky, L.~Frankfurt, J.F.~Gunion, A.H.~Mueller, and M.~Strikman,
Phys.\ Rev.\ D {\bf 50}, 3134 (1994);
A.V.~Radyushkin,
Phys.\ Lett.\ B {\bf 385}, 333 (1996);
L.~Mankiewicz, G.~Piller, and T.~Weigl, Eur.\ Phys.\ J.\ C {\bf 5}, 119 (1998);
Phys.\ Rev.\ D {\bf 59}, 017501 (1999);
M.~Vanderhaeghen, P.A.M.~Guichon, and M.~Guidal,
Phys.\ Rev.\ Lett.\ {\bf 80}, 5064 (1998);
\bibitem{maxben}
B.~Lehmann-Dronke, P.V.~Pobylitsa, M.V.~Polyakov, A.~Sch\"afer, and
K.~Goeke, Phys.\ Lett.\ B {\bf 475}, 147 (2000);
B.~Lehmann-Dronke, M.V.~Polyakov, A.~Sch\"afer, and K.~Goeke,
Phys.\ Rev.\ D {\bf 63}, 114001 (2001).
\bibitem{nufact}
B.J.~King, hep-ex/9907033;
hep-ex/9911008;
Nucl.\ Instr.\ Meth.\ A {\bf 451}, 198 (2000).
\bibitem{nudis}
M.L.~Mangano et~al., hep-ph/0105155.
\bibitem{csprod}
A.E.~Asratyan et~al., Bigg Bubble Chamber Neutrino Collaboration,
Z.\ Phys.\ C {\bf 58}, 55 (1993);
P.~Annis~et al., CHORUS Collaboration, Phys.\ Lett.\ B {\bf 435}, 458 (1998).
\bibitem{Ji}
X.~Ji, J.\ Phys.\ G {\bf 24}, 1181 (1998).
\bibitem{Bauer}
M.~Wirbel, B.~Stech, and M.~Bauer, Z.\ Phys.\ C {\bf 29}, 637 (1985);
M.~Bauer and M.~Wirbel, Z.\ Phys.\ C {\bf 42}, 671 (1989).
\bibitem{omega}
H-n.~Li and B.~Meli\'c, Eur.\ Phys.\ J.\ C {\bf 11}, 695 (1999).
\bibitem{fDs}
A.~Abada et~al., Nucl.\ Phys.\ Suppl. {\bf 83},268 (2000);
D.~Becirevic et~al., hep-lat/0002025;
A.~Ali Khan et~al., CP-PACS Collaboration,
Phys.\ Rev.\ D {\bf 64}, 034505 (2001);
A.S.~Kronfeld, hep-ph/0010074;
L.~Lellouch and C.J.D.~Lin, UKQCD Collaboration, hep-ph/0011086.
\bibitem{Rich}
J.D.~Richman and P.R.~Burchat, Rev.\ Mod.\ Phys. {\bf 67}, 893 (1993).
\bibitem{Rad99}
A.V.~Radyushkin, Phys.\ Rev.\ D {\bf 59}, 014030 (1999).
\bibitem{MRS}
A.D.~Martin, R.G.~Roberts, and W.J.~Stirling,
Phys.\ Lett.\ B {\bf 354}, 155 (1995).
\bibitem{GS}
T.~Gehrmann and W.J.~Stirling,
Phys.\ Rev.\ D {\bf 53}, 6100 (1996).
\bibitem{BGMS93}
V.M.~Braun, P.~Gornicki, L.~Mankiewicz, and A.~Sch\"afer,
Phys.\ Lett.\ B {\bf 302}, 291 (1993).
\bibitem{fDsexp}
S.~Aoki et~al., WA75 Collaboration, Prog.\ Theor.\ Phys. {\bf 89}, 695 (1993);
D.~Acosta et~al., CLEO Collaboration, Phys.\ Rev.\ D {\bf 49}, 5690 (1994);
K.~Kodama et~al., Fermilab E653 Collaboration, Phys.\ Lett.\ B
{\bf 382}, 299 (1996);
M.~Acciarri et~al., L3 Collaboration, Phys.\ Lett.\ B {\bf 396}, 327 (1997);
J.Z.~Bai et~al., BES Collaboration, Phys.\ Rev.\ D {\bf 57}, 28 (1998);
M.~Chada et~al., CLEO Collaboration, Phys.\ Rev.\ D {\bf 58}, 032002 (1998);
Y.~Alexandrov et~al., BEATRICE Collaboration, Phys.\ Lett.\ B
{\bf 478}, 31 (2000).
\bibitem{Migliozzi}
G.~De Lellis, P.~Migliozzi, and P.~Zucchelli, Phys.\ Lett.\ B
{\bf 507}, 7 (2001).
\end{thebibliography}
\end{document}